\documentclass[aps,showpacs,preprintnumbers,amsmath,amssymb]{revtex4}

\usepackage{float}
\usepackage{graphics,epsfig}
\usepackage{graphicx}
\usepackage{dcolumn}
\usepackage{bm}

\begin{document}
\baselineskip=0.8 cm
\title{{\bf Signature of the black hole phase transition in quasinormal modes}}

\author{Xi He, Bin Wang}
\affiliation{Department of Physics, Fudan University, 200433 Shanghai}
\author{Rong-Gen Cai}
\affiliation{Institute of Theoretical Physics, Chinese Academy of Sciences, P.O. Box 2735, 100190 Beijing}
\author{Chi-Yong Lin}
\affiliation{Department of Physics, National Dong Hwa University, Shoufeng, 974 Hualien}

\vspace*{0.2cm}
\begin{abstract}
\baselineskip=0.6 cm
\begin{center}
{\bf Abstract}
\end{center}

We study the perturbation of the scalar field interacting with the Maxwell field in the background of d-dimensional charged AdS black hole and
AdS soliton.  Different from the single classical field perturbation, which always has the decay mode in the black hole background, we observe
the possible growing mode when the perturbation of the scalar field strongly couples to the Maxwell field. Our results disclose the signature of
how the phase transition happens when the interaction among classical fields is strong. The sudden change of the perturbation to growing mode is
also observed in the AdS soliton with electric potential. However in the magnetic charged AdS soliton background, we observe the consistent
perturbation behavior when the interaction between scalar field and Maxwell field is considered. This implies that for the magnetic charged AdS
soliton configuration, unlike the situation with electric potential, there is no scalar field condensation which causes the phase change.

\end{abstract}

\pacs{ 04.70.Dy, 95.30.Sf, 97.60.Lf } \maketitle
\newpage
\section{introduction}

The quasinormal mode (QNM) of black holes describes the damped oscillations under perturbations in the surrounding geometry of a black hole with
frequencies and damping times of the oscillations entirely fixed by the black hole parameters. The QNM is considered as the characteristic sound
of black holes and is expected to be detected through gravitational wave observations \cite{1,2}. In addition to its potential astrophysical
interest to directly identify the black hole existence, the black hole QNM has been argued recently as a useful testing ground for fundamental
physics. It is widely believed that the study of QNM can help us get deeper understandings of the AdS/CFT \cite{3, 4,41,42,43}, dS/CFT \cite{5}
correspondences \cite{33}, loop quantum gravity \cite{6} and present us the imprint of the extra dimensions \cite{7,8}. Furthermore it has been
argued that the QNM can reflect the black hole phase transition \cite{9,10,11}.

In most studies of the QNM, the wave dynamics of a single classical field propagating outside the black hole is taken into account. For the
stable black hole backgrounds, the perturbation of a single classical field  will die out finally due to the black hole absorption. Recently more
interest has been focused on the strongly coupled theory by considering the Einstein-Maxwell field interacting with a charged scalar field with a
minimal Lagrangian
\begin{eqnarray}\label{Lag}
\mathcal{L}=-{1\over 4} F^2_{\mu\nu}-|\partial_\mu \Psi-iqA_\mu \Psi|^2-m^2|\Psi|^2.
\end{eqnarray}
In the light of AdS/CFT correspondence, it was realized that a condensate of the charged scalar field is formed through its coupling to a Maxwell
field of the background \cite{12}. Along this line, there have been a lot of investigations concerning the application of AdS/CFT correspondence
to condensed matter physics by considering interactions among classical fields \cite{13,14,15,16,17,18,19}. See Refs. \cite{20, 21} for reviews.
It was disclosed that there is a second order transition, with mean field theory exponents, between a non-superconducting state at high
temperatures, where all the charge is in the normal component, and a superconducting state at low temperatures. Similar transition occurs for
black hole geometries based on gravity coupled to the Abelian Higgs model \cite{23,25}. It is of great interest to ask whether the QNM can be an
effective probe of this phase transition. This is the motivation of the present work.

In \cite{Gubser}, the marginally stable linearized perturbation
around a charged black hole background in AdS space that breaks the
$U(1)$ symmetry has been discussed. In this work we want to
establish the more range of behaviors that this mechanism exhibits
by studying the QNM behavior. In \cite{30}, the QNM at strong
coupling was computed from the poles of retarded Green's function in
the black hole background. In our investigation we will start from
the bulk equations and directly study the wave dynamics outside
black holes. By this way we can get the objective picture of how the
perturbation of the scalar field evolves if it is coupled to a
Maxwell field. By comparing our result with that obtained in
\cite{30}, we can have further insight into the holographic
correspondence.

Besides the black hole configuration, in our study we will also investigate the AdS soliton configuration which has lower energy than the AdS
space in the Poincare coordinates, but has the same boundary topology as the Ricci flat black hole and the AdS space in the Poincare coordinates
\cite{224}. It was found that there is a Hawking-Page phase transition between the Ricci flat AdS black hole and the AdS soliton \cite{223,225}
and this  phase transition was observed in the QNM spectrum \cite{10}. Recently it was argued that a  phase transition between an AdS soliton
configuration and a superconducting phase will appear if the chemical potential is changed \cite{227}. We are going to investigate whether this
phenomenon can be reflected in the QNM behavior.

The organization of the paper is as follows. In Sec.~II we will
present the charged black hole backgrounds of interest and discuss
associated  wave dynamics . In Sec.~III we will investigate the QNMs
of AdS soliton. Conclusions and discussions will be presented in
Sec.~IV.

\section{The Reissner-Nordstr\"{o}m (RN) black holes in AdS spaces}

The electrically charged black hole in $d$-dimensional AdS space is described by the metric
\begin{eqnarray}\label{RN}
ds^2=-f dt^2+ {1\over f}dr^2 +r^2 h_{ij} dx^i dx^j,
\end{eqnarray}
where
\begin{eqnarray}\label{f}
f=k-{2M\over r^{d-3}}+{Q^2\over 4 r^{2d-6}}+{r^2\over L^2},
\end{eqnarray}
and $M$ and $Q$ are referred to black hole mass and charge. $L$ here
is the AdS radius. The black hole horizon $r_+$ can be obtained from
$f(r)=0$. $h_{ij} dx^i dx^j$ is the line element for flat ($k=0$),
spherical ($k>0$) or hyperbolic ($k<0$) $(d-2
)$-plane. The Hawking
temperature defined as $T=f'(r_+)/4\pi$ is
\begin{eqnarray}
T={r_+^{-2d-1}\over 16 \pi
L^2}\bigg[4\big(d-1\big)r_+^{2d+2}-\big(d-3\big)L^2\big(Q^2r_+^6-4kr_+^{2d}\big)\bigg].
\end{eqnarray}

The gauge field ansatz is
\begin{eqnarray}\label{gbh}
A=\Phi(r)dt.
\end{eqnarray}
We can find a simple solution to the equations of motion from the Lagrangian (\ref{Lag}) together with the Einstein-Hilbert Lagrangian plus a
negative cosmological constant, which is in the form
\begin{eqnarray}
\Phi=\sqrt{d-2\over 2(d-3)}\bigg({Q\over r^{d-3}}-{Q\over
r_+^{d-3}}\bigg),~~~~~~~ \Psi=0.
\end{eqnarray}

Considering the scalar field $\Psi$ perturbing around the RN-AdS$_d$ black hole, we have the wave equation of the charged scalar field directly
from the Lagrangian (\ref{Lag})
\begin{eqnarray}\label{aa}
\bigg[{1\over \sqrt{-g}}{\partial \over \partial
x^\mu}\bigg(\sqrt{-g}g^{\mu \nu}{\partial \over
\partial x^\nu}\bigg)-m_{eff}^2\bigg]\Psi=0,
\end{eqnarray}
where the effective mass of the scalar field is of the form
\begin{eqnarray}\label{meff}
m_{eff}^2=m^2+g^{tt}q^2\Phi^2.
\end{eqnarray}
Since $g^{tt}$ is negative outside the horizon and diverges to $-\infty$ near the horizon, it seems possible that $m_{eff}^2$ should become
negative which will directly lead to the unstable perturbation in the following discussion. The radial part of the equation (\ref{aa}) can be
separated by setting $\Psi=e^{-i\omega t}{r^{2-d\over 2}}R(r)S(x_i)$. Defining the tortoise coordinate $dr_*= dr/f(r)$, the radial wave equation
can be expressed as
\begin{eqnarray}\label{wave}
{d^2R(r) \over dr_*^2}+\bigg[\omega^2-V(r)\bigg]R(r)=0,
\end{eqnarray}
with the effective potential
\begin{eqnarray}
V(r)={(d-2)(d-4)\over {4 r^2}}f^2 +{\lambda^2 \over r^2}f +m_{eff}^2 f +{d-2\over 2r}ff',
\end{eqnarray}
and the separation constant $\lambda^2=l(l+d-3)$, ($l=0,1,2...$). Throughout this paper we will only consider $l=0$ for simplicity.

\begin{figure}
\includegraphics[width=6cm]{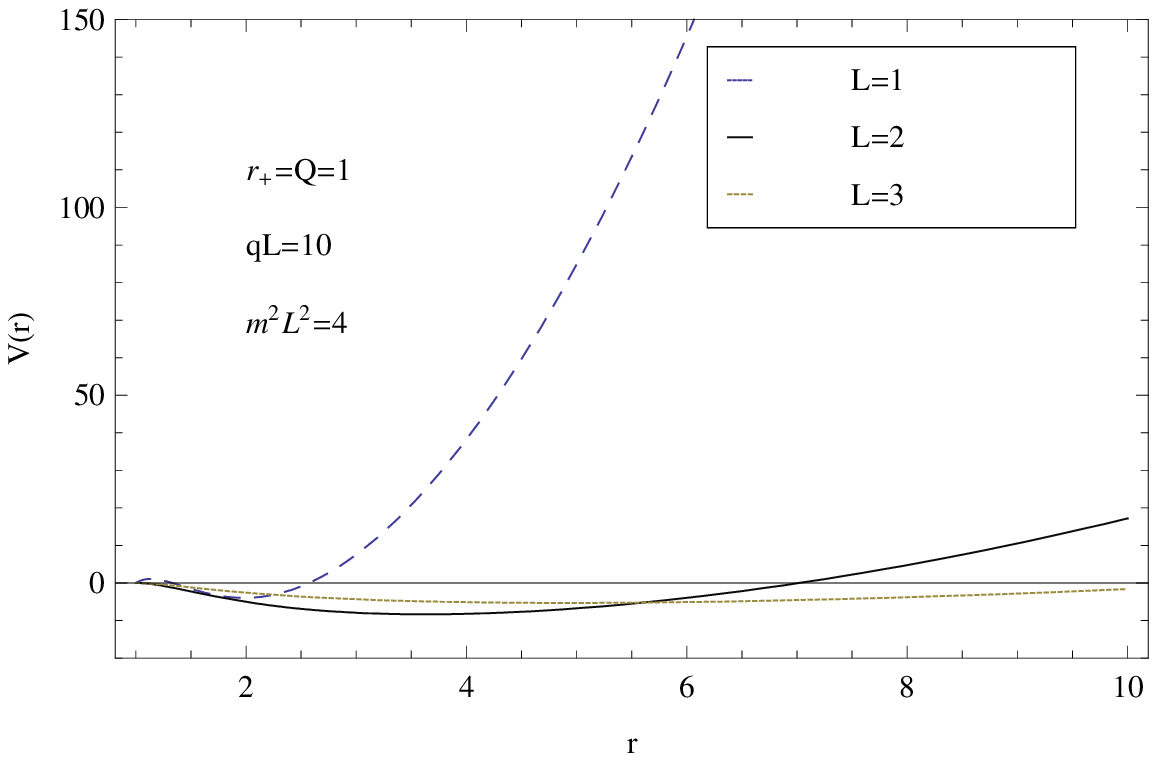}
\includegraphics[width=6cm]{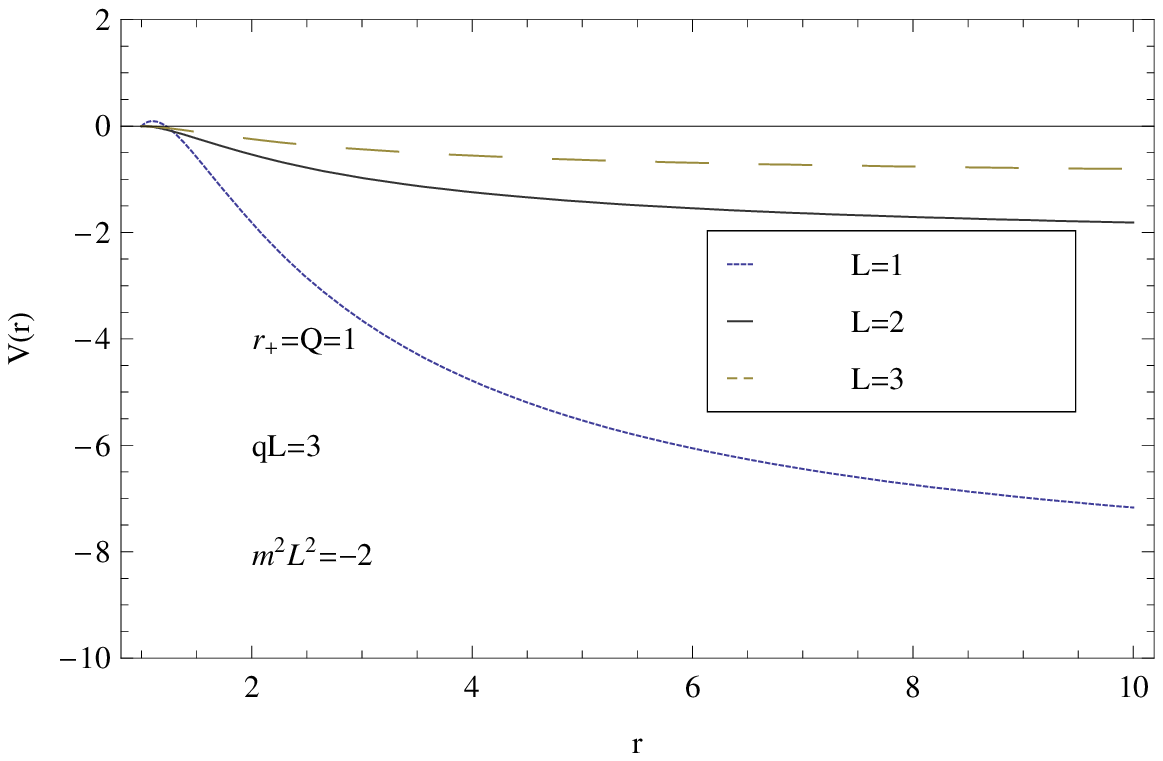}
\caption{\label{V} The behavior of the effective potential.  The left panel is for selecting $qL=10$, $m^2 L^2=4$, and the right panel is for
$qL=3$, $m^2 L^2=-2$. In both cases we set $r_+=Q=1$.}
\end{figure}

Choosing scaling symmetry adopted in \cite{Gubser}, we can fix
$r_+=Q=1$ and change the value of $L$ to vary the temperature of the
black hole. We follow \cite{Gubser} to fix $m^2L^2$ and $qL$ as we
vary $L$. For selecting $m^2L^2=4, ~qL=10$, the effective potential
for the RN-AdS$_4$ black hole background is shown in the left of
Fig.\ref{V}. The negative potential well appears due to the negative
value of the effective mass term in Eq.(\ref{meff}). When
$r\rightarrow \infty$, the potential approaches the positive
infinity. With the increase of the AdS radius, the temperature of
the black hole becomes smaller and we see that the negative
potential well becomes wider. The behavior of the effective
potential changes drastically when $m^2<0$. Choosing $m^2L^2=-2,
~qL=3$, we show the effective potential in the right of Fig.\ref{V}.
We see that the potential is almost negative except a small positive
barrier near horizon for small $L$. When $r\rightarrow \infty$, the
potential approaches $V(r)\rightarrow -9/L^2$. For other dimensions,
the potential behaviors are similar. It was argued in \cite{5} that
the behavior of the effective potential influences a lot on the wave
dynamics. In the following we are going to examine the evolution of
the wave propagation of the perturbed field.

Using the new variables $v=t-r_*$, $u=t+r_*$, we can rewrite Eq.(\ref{wave}) into
\begin{eqnarray}
{d^2R \over du dv} + {V(r)\over 4} R=0.
\end{eqnarray}
This equation can be discretized as
\begin{eqnarray}
R_N=R_E+R_W-R_S+\triangle u \triangle v V\bigg({v_N+v_W-u_N-u_E\over 4}\bigg){R_W+R_E\over 8}+O(\epsilon^2),
\end{eqnarray}
where the points $N$, $S$, $E$, and $W$ form a null rectangle with relative positions as: $N: (u+\triangle u,~v+\triangle v)$,  $W: (u+\triangle
u,~v)$, $E: (u,~v+\triangle v)$ and $S: (u,~v)$. The parameter $\epsilon$ is an overall grid scale factor, so that $\triangle u \sim \triangle v
\sim \epsilon$.

In presenting the numerical results, we fix $r_+=Q=1$ and vary the
AdS radius $L$ which corresponds to the change of the black hole
temperature. The left panel of Fig.2 shows the objective picture of
the evolution of perturbation for choosing $m^2L^2=4, ~qL=10$ in the
4-dimensional flat black hole background. We observe that with the
increase of $L$, the temperature of the black hole drops, the
perturbation oscillates less and lasts longer. When $L_*=1.11$, the
perturbation does not oscillate and stops decaying. This value of
$L$ coincides with the marginally stable mode observed in
\cite{Gubser}. When $L>L_*$, the perturbation outside the black hole
keeps growing. The blow up of the perturbation indicates that when
the black hole temperature is under some critical value, the
original black hole configuration is no longer stable. When this
occurs, the scalar will start to condensate and address to the black
hole configuration, which was described by the superconducting state
in the language of AdS/CFT correspondence \cite{20}.

\begin{figure}
\includegraphics[width=14cm]{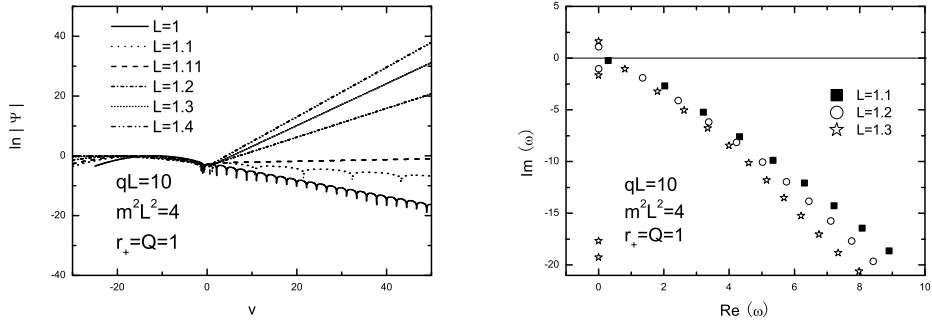}
\caption{\label{1a} The QNM of charged scalar field around flat RN-AdS$_4$ black hole when we take $qL=10$, $m^2L^2=4$.  }
\end{figure}

\begin{figure}
\includegraphics[width=14cm]{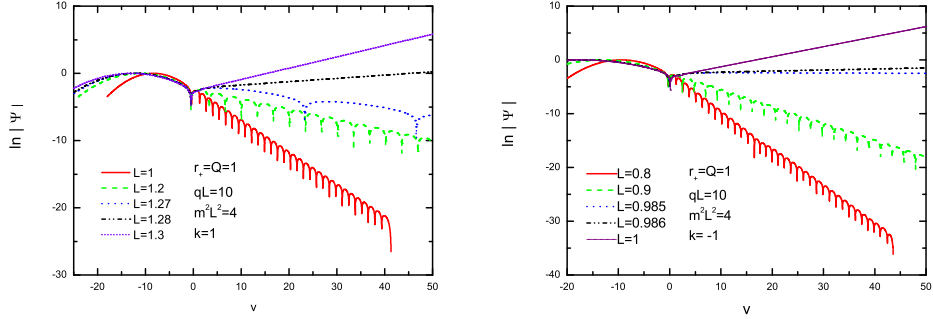}
\caption{\label{1a} The QNM of charged scalar field around the spherical RN-$AdS_4$ black hole (left) and hyperbolic RN-$AdS_4$ black hole
(right) for setting $qL=10$, $m^2L^2=4$.  }
\end{figure}

\begin{figure}
\includegraphics[width=14cm]{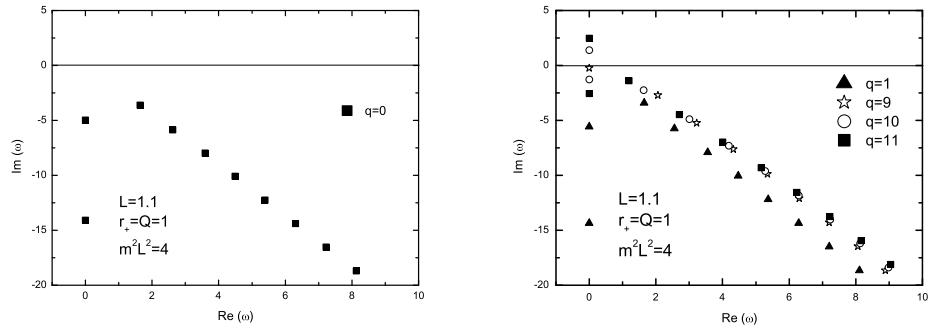}
\caption{\label{2a} The influence of the quasinormal frequency in the background of the flat RN-AdS$_4$ black hole due to the value of $q$. Here
we set $L=1.1$, $m^2L^2=4$. The left panel is for $q=0$ and the right one is for the increase of $q$. }
\end{figure}

\begin{figure}
\includegraphics[width=14cm]{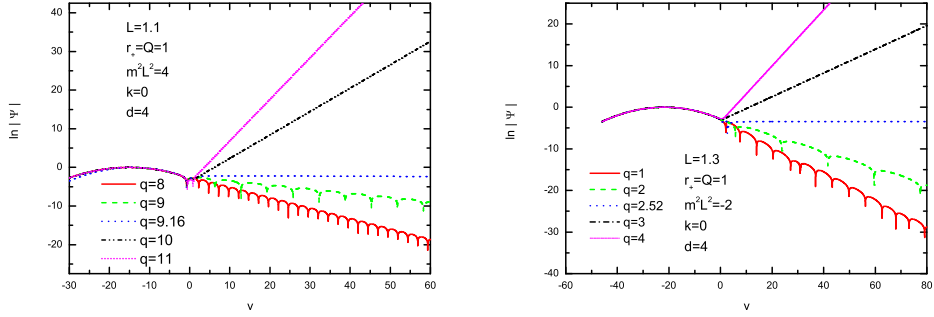}
\caption{\label{3} The objective picture of the dependence of the QNM of flat RN-AdS$_4$ black hole on values of $q$. The left panel is for
choosing $m^2L^2=4$, $L=1.1$ and we find $q_*=9.16$, while  the right panel is for selecting $m^2L^2=-2$, $L=1.3$ and we have $q_*=2.52$.}
\end{figure}

\begin{figure}
\includegraphics[width=7cm]{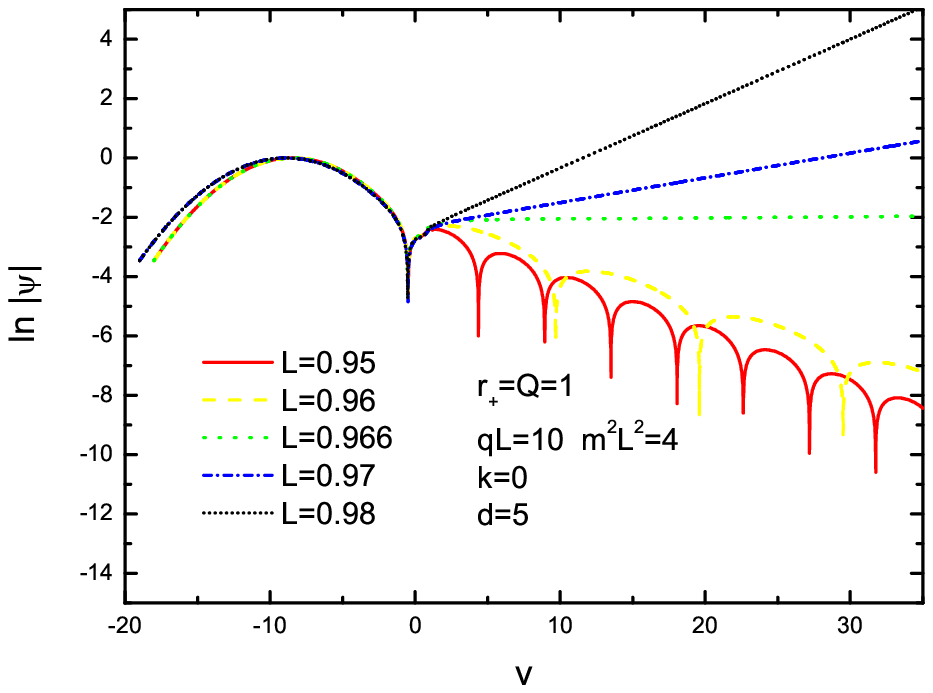}
\includegraphics[width=7cm]{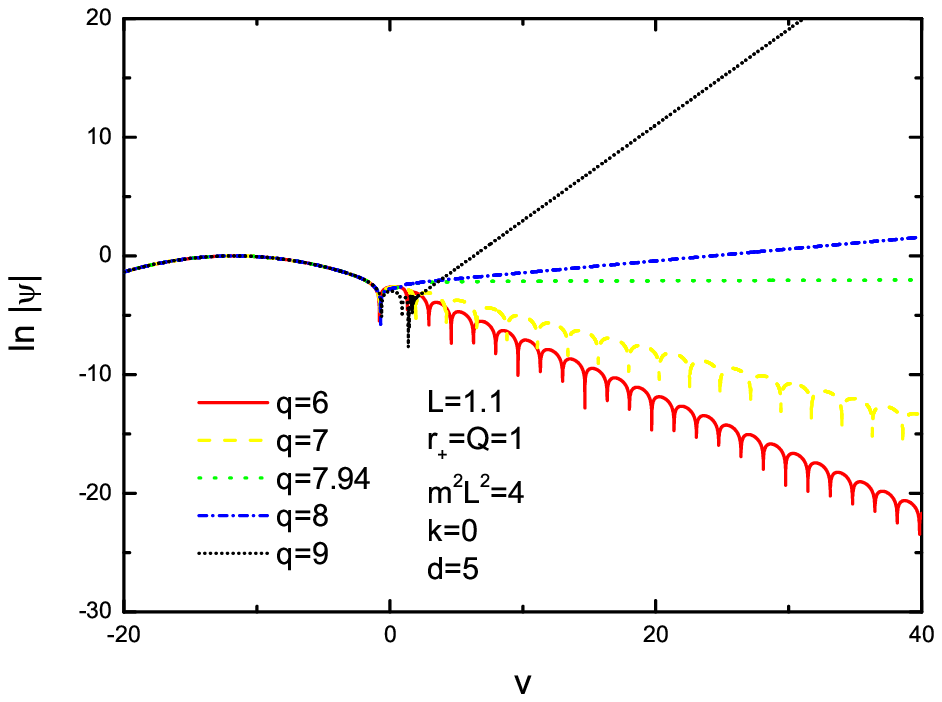}
\caption{\label{newd5} The objective picture of the QNM for
RN-AdS$_5$ black hole. The left panel is for selecting $qL=10$, $m^2
L^2=4$ and we get the critical AdS radius $L_*=0.966$. The right
panel shows the result changing with $q$ by fixing $L=1.1$, $m^2
L^2=4$ and we get the critical value of the charge of the scalar
field to break the stability at $q_*=7.94$. }
\end{figure}

Considering the case for $m^2L^2=4, ~qL=10$, the potential
approaches positive infinity, which puts $\Psi=0$ at the boundary.
We can use the Horowitz-Hubeny method \cite{3}, which has been used
extensively in previous papers \cite{4,41} to directly calculate the
quasinormal frequencies of the perturbation. The result is presented
in the right panel of Fig.2. As the increase of $L$, the black hole
temperature drops, the topmost mode will move into the upper half
frequency plane, indicating the growing mode and the instability of
the black hole configuration. The critical value $L_*=1.11$ agrees
with that observed in the objective picture. The behavior of the
quasinormal frequency we obtained from solving the bulk equation is
in consistent with that from poles in the retarded Green's function
\cite{30}.

In Fig.3 we also present the result of wave dynamics for 4-dimensional spherical and hyperbolic black hole backgrounds. Choosing the same values
of $m^2L^2, qL$ as those of the flat case, we see that in the spherical black hole background, the black hole starts to be unstable due to the
perturbation at bigger value of the AdS radius $L_*$, while for the hyperbolic case, the black hole begins unstable at smaller $L_*$ than that of
the flat case. Inserting these $L_*$ into the temperature expression, we observe that the critical temperature to break the background stability
satisfies $T_{*(k=1)}>T_{*(k=0)}>T_{*(k=-1)}$. This shows that the topology of the spacetime has the influence on evolution of the perturbation.
The spherical background is the easiest to be destroyed due to the perturbation. This phenomenon provides the motivation to examine the spacial
topology influence on the condensation. Recently it was argued that the negative curvature topology can make the superconductor gapless and give
a geometrical mechanism of conductivity \cite{17}. More studies in this direction is called for.

The growing mode was caused by the interaction between scalar field and the Maxwell field, which has never been observed in the perturbation of a
single classical field in the black hole background \cite{4,41,42}. The process of the blow up (or the scalar condensation) can be understood as
arising in part from the temperature of the black hole, but also as arising from the other two factors which lead to the negative effective mass
of the scalar field, the charge $q$ and the mass $m$ of the scalar field. To see the influence of these two factors, we fix the value of the
black hole background, $r_+, Q$ and $L$. If $q$ is big, the electric field outside the horizon is big. For the same mass of the scalar field,
$m^2_{eff}$ will become negative provided that $q$ is big enough, this can be seen from (\ref{meff}). The influence of $q$ on the evolution of
the perturbation is shown in Fig.4. When $q=0$, there is no coupling between the scalar field and the Maxwell field, the perturbation always
decays. However with the increase of the $q$, it is possible for $m^2_{eff}$ to become negative. The topmost mode moves upward with the increase
of $q$ and when $q$ is over some critical value $q_*$ the topmost mode will move into the upper half of the frequency plane indicating the
instability of the background. The objective picture with the increase of $q$ is also shown in Fig.5. For smaller mass of the scalar field, it is
easier for $m^2_{eff}$ to be negative outside the horizon and to produce an unstable mode in $\Psi$, this can happen even for the smaller value
of $q_*$, which is shown in the right panel of Fig.5.

In Fig.6, we also exhibit the result of the perturbation for
5-dimensional flat black hole background. Comparing with that of the
4-dimensional situation (shown in Fig.2), the blow up of the
perturbation can appear at smaller values of $L_*$ and $q_*$.  For
fixing $r_+=Q=1$, inserting the values of $L_*$ into the temperature
expression, we have $T_{*4D}>T_{*5D}$. This shows that the
instability of the 4-dimensional RN AdS black hole can happen at
higher temperature than that in 5-dimensions.  However when we fix
the same AdS radius in 4-dimensions and 5-dimensions, we observe
that the critical value of the charge of the scalar field to break
the stability is smaller for the 5-dimensional case. This implies
that for the same black hole parameters ($r_+, Q, L$), the stability
of the black hole spacetime can be broken easier in the high
dimensions. This result supports the finding that the scalar hair
can be formed easier in the higher dimensional background \cite{18}.

\section{The AdS soliton background}

In this section we will study the perturbation of the scalar field coupled to Maxwell field in the background of AdS soliton. Recently the
signature of the Hawking-Page phase transition between the Ricci flat AdS black hole and the AdS soliton \cite{224,223,225} was observed in the
QNM spectrum of the scalar perturbation in \cite{10}. It was argued that there is a phase transition in the AdS soliton configuration and the
superconducting phase will appear dual to an AdS soliton \cite{227}. It is of interest to study the evolution of the perturbation in the charged
AdS soliton background and get more understanding on this phase transition.

By analytically continuing the Ricci flat black hole one obtains the AdS soliton \cite{224,223}
\begin{eqnarray}\label{sol}
ds^2=-r^2 dt_s^2+{1\over f_s}dr^2  +f_s d\phi_s^2+r^2 h_{ij}dx^i dx^j.
\end{eqnarray}
To satisfy the Einstein equation the function $f_s$ is in the same form as Eq.(\ref{f}) but with a negative sign before $Q^2$. We have  $k=0$ and
$r_{s}$ replacing $r_+$,  $\phi_s$ is identified as the period $\beta_s= 4 \pi / f'(r_{s})$ and $h_{ij}dx^idx^j$ denotes a $(d-3)$-dimensional
Euclidean space. Now the the gauge field is transformed into
\begin{eqnarray}\label{gs}
A=\Phi(r)d\phi_s.
\end{eqnarray}
This corresponds to a magnetic field $F_{\phi_s r}=d\Phi(r)/dr$, while Eq.~(5) leads to a radial electric field.

With the variable separation $\Psi(r)=e^{-i\omega t}  {r^{{2-d}\over 2}}R(r) S(\theta) Y(\phi_s)$, we obtain
\begin{eqnarray}
{d^2R(r)\over dr^2}+{f_s'\over f_s}{d R(r)\over dr}+ \bigg[-{(d-2)(d-4)\over 4r^2}- {d-2\over 2r}{f'\over f}+{\omega^2 \over
fr^2}-{m_{eff}^2\over f}-{\lambda^2 \over fr^2}-{k_s^2\over f^2}\bigg]R(r)=0,
\end{eqnarray}
where the effective mass is
\begin{eqnarray}
m_{eff}^2=m^2+g^{\phi \phi}q^2 \Phi^2,~~~~ \Phi = \sqrt{{d-2\over 2(d-3)}}\bigg({Q\over r_s}-{Q\over r}+A\bigg).
\end{eqnarray}
$A$ here is a constant. $\lambda^2$ and $k_s^2={2\pi \tilde{m}\over \eta_s}$ ($\tilde{m}=0,1,2...$) are the eigenvalues of
\begin{eqnarray}
&&{d^2S(\theta)\over d \theta^2}+\lambda^2 S(\theta)=0, \\
\nonumber &&{d^2Y(\phi_s)\over d \phi_s^2}+k_s^2 Y (\phi_s)=0.
\end{eqnarray}

We observed that the negative effective mass term plays an important
role in producing growing modes in the RN-AdS black hole. However in
the charged AdS soliton case, due to the coordinate transformation,
$g^{\phi \phi}$ substitutes $g^{tt}$ in the second term of the
effective mass. Since $g^{\phi \phi}$ is always positive, the
effective mass will always be positive provided that $m^2$ is not
too negative. This leads to the consistent result on the QNMs of AdS
soliton as disclosed in \cite{10} that only normal modes exist as
exhibited in table I and table II. Note that here the numerical
results are shown in the case of $\lambda=k_s=0$.

\begin{table}
\caption{\label{TableSN1}Normal Modes for magnetic charged AdS$_4$ soliton, we set $r_+=Q=L=1$, $q=10$. We show the results of the first three
modes with overtone numbers $n=0,1,2$ from the top to the bottom.}
\begin{center}
\begin{tabular}{|c|c|c|c|}
\hline
$m^2=2$&$m^2=1$ & $m^2=0 $& $m^2=-1$\\
\hline
$4.0612862$&$3.8935134$&$3.7161216 $&$3.5271927$\\
$6.7874836$&$6.6222613$&$6.4452027 $&$6.2519047$\\
$9.0955200$&$8.9047114$&$8.6926610 $&$8.4464558$\\
\hline
\end{tabular}
\end{center}
\end{table}

\begin{table}
\caption{\label{TableSN2}Normal Modes for magnetic charged AdS$_4$ soliton, we set $r_+=Q=L=1$, $m^2=0$ . We show the results of the first three
modes with overtone numbers $n=0,1,2$ from the top to the bottom.}
\begin{center}
\begin{tabular}{|c|c|c|c|}
\hline
$q=10$ & $q=5 $& $q=1$& $q=0$\\
\hline
$3.7161216$&$3.0973695 $&$2.7779244$&$2.7621022$\\
$6.4452027$&$5.4999791 $&$5.1024004$&$5.0846170$\\
$8.6926610$&$7.7412947 $&$7.3972210$&$7.3825090$\\
\hline
\end{tabular}
\end{center}
\end{table}

This result indicates that the behavior of the perturbation of the scalar field coupling to Maxwell field does not change in the AdS soliton
background with magnetic charged field, which is different from that in the charged AdS black hole. This property holds for higher dimensions.
Unlike the charged AdS black hole that the perturbation will blow up when the black hole temperature is low enough or the coupling between scalar
field and Maxwell field is strong enough, the perturbation in magnetic charged AdS soliton always keeps the normal modes. At the first sight,
this is in contrary with the result reported in \cite{227}, where it was argued that like the AdS black hole case, there is holographic
superconductor phase where the charged scalar condenses in the AdS soliton configuration. However in their discussion, they started with the
neutral AdS soliton with the electric field where the electric potential is constant and the field strength is zero. In doing so $g^{tt}$ will
appear in the effective mass term in their discussion. However in the magnetic charged AdS soliton background, the coordinate transformation
forces $A=\Phi(r)d\phi_s$ to satisfy the Einstein equation and this brings $g^{\phi \phi}$ in the effective mass instead of the magic term
$g^{tt}$ which can make the effective mass negative and lead to the drastic change in the behavior of the perturbation. Thus for the magnetic
charged AdS soliton background, the consistent perturbation behavior tells us that there is no drastic phase change and the scalar field may not
condensate in this configuration. This result is different from that of the neutral AdS soliton with the electric field. For the comparison we
present numerical results of the perturbation in the neutral AdS soliton with the electric field in table III, where we find that when the
electric potential is bigger, we do see that the quasinormal frequencies suddenly change from the purely real value to purely imaginary. The
critical value of the electric potential to allow the sudden change of the quasinormal frequency decreases with the decrease of the mass of the
scalar field, i.e. $\Phi_{critical}=1$ for $m^2=0$, $\Phi_{critical}=2$ for $m^2=1$ and $\Phi_{critical}=3$ for $m^2=4$. The positive imaginary
frequency indicates that when the coupling between scalar and Maxwell field is strong enough, the AdS soliton with the electric field can become
unstable and the superconducting phase can be formed.

\begin{table}
\caption{\label{TableSN3}Perturbations for the neutral AdS soliton with the electric field, we set $r_+=L=1$, $m^2=4$ . We show the results of
the first three modes with overtone numbers $n=0,1,2$ from the top to the bottom.}
\begin{center}
\begin{tabular}{|c|c|c|c|}
\hline
$\Phi=0$ & $\Phi=2 $& $\Phi=3$& $\Phi=4$\\
\hline
$2.7458368$&$1.8813878 $&$1.2084619 i$&$2.9086733 i$\\
$3.9089738$&$3.3585824 $&$2.5060080$&$0.8484833 i$\\
$4.9738799$&$4.5540620 $&$3.9673015$&$2.9562613$\\
\hline
\end{tabular}
\end{center}
\end{table}

\section{conclusions and discussions}

In this work we have studied the QNM of the perturbation of the scalar field interacting with the Maxwell field. Different from the single
classical field perturbation, which always has the decay mode in the black hole background, we observed the possible growing mode when the
perturbation of the scalar field strongly couples to the Maxwell field. This indicates that the interaction among classical fields can destroy
the background spacetime. In the language of the AdS/CFT, this is due to the condensate of the scalar field on the black hole background and
there is a second order transition between a non-superconducting state at high temperatures and a superconducting state at low temperatures. Our
results disclose the signature of how this phase transition happens from the phenomenon of the perturbation. In the black hole background we have
seen the influence of the topology and the dimensionality on the QNM when the scalar field is coupled with the Maxwell field.

We also discussed the perturbation in the AdS soliton background with magnetic field. We observed the consistent normal mode of the perturbation
in the AdS soliton background with magnetic field and there is no characteristic of the phase transition in this configuration. This is different
from that of the RN-AdS black hole background and indicates that possibly there is no scalar condensation in this AdS soliton configuration. In
the discussion of [33], they started with the neutral spacetime with the electric field where the electric potential is constant and the field
strength is zero. We have shown the signature of how the phase transition happens from the phenomenon of the perturbation in this background. It
would be more interesting to investigate the electric charged AdS soliton background. For the AdS soliton background with electric charge, the
exact solution of the Einstein equation can be found in the form
\begin{eqnarray}
ds^2 = fdz^2 +f^{-1}dr^2 +(-r^2 dt^2 +r^2 dy^2)
\end{eqnarray}
where $f = -m/r -q^2/r^2 +r^2/l^2$, and $A= qydt$ is the gauge field. The appearance of  the $y$ coordinate in the electric field makes the
perturbation calculation much complicated. This solution itself is a new one and we will study this new soliton solution carefully in a separate
future work.

Our investigation is still limited in the probe limit and has not scanned the full parameter space in the wave equations. Further work is needed
to establish the full range of behaviors that the mechanism of the perturbation exhibits. Besides it is also of interest to include the usual
$|\Psi|^4$ term in the lagrangian and disclose its scalar perturbation dynamics, which will help get deeper understanding on the holographic
condensation by comparing with the usual Ginzburg-Landau story. We look forward to reporting some of the progress on these issues in the future.

\begin{acknowledgments}
This work was partially supported by NNSF of China. We would like to acknowledge helpful discussions with Q. Y. Pan, S. Y. Yin, R. K. Su, S. F.
Wu, X. H. Ge and Y. Q. Liu.

\end{acknowledgments}

\end{document}